\documentclass[
reprint,
superscriptaddress,
groupedaddress,
amsmath,amssymb,
aps,
prb,
]{revtex4-2}

\usepackage{graphicx}
\usepackage{dcolumn}
\usepackage{bm}
\usepackage{hyperref}
\usepackage[dvipsnames]{xcolor}
\usepackage{soul}
\usepackage[left]{lineno}

\begin{document}

\title{Many-body origin of anomalous Floquet phases in cavity-QED materials}
\author{Beatriz Pérez-González}
\email{bperez03@ucm.es}
\author{Gloria Platero}
\email{gplatero@icmm.csic.es}
\affiliation{Instituto de Ciencia de Materiales de Madrid (ICMM), CSIC, Calle Sor Juana Inés de la Cruz 3, 28049 Madrid, Spain}
\author{\'Alvaro~G\'omez-Le\'on}
\email{a.gomez.leon@csic.es}
\affiliation{Institute of Fundamental Physics IFF-CSIC, Calle Serrano 113b, 28006 Madrid, Spain}
\date{\today}
\begin{abstract}
Anomalous Floquet topological phases are a hallmark, without a static analog, of periodically driven systems. Recently, Quantum Floquet Engineering has emerged as an interesting approach to cavity-QED materials, which recovers the physics of Floquet engineering in its semi-classical limit. However, the mapping between these two widely different scenarios remains mysterious in many aspects.
We discuss the emergence of anomalous topological phases in cavity-QED materials, and link topological phase transitions in the many-body spectrum with those in the $0$- and $\pi$-gaps of Floquet quasienergies. Our results allow to establish the microscopic origin of an emergent discrete time-translation symmetry in the matter sector, and link the physics of isolated many-body systems with that of periodically driven ones. Finally, the relation between many-body and Floquet topological invariants is discussed, as well as the bulk-edge correspondence.
\end{abstract}
\maketitle
\section{Introduction}
Topological systems in condensed matter have attracted wide attention during the last decades. From a theory perspective they exhibit robust and exotic physical properties which can be characterized using simple topological arguments only~\cite{Laughlin1981,Thouless1982}.
From an experimental point of view they are very interesting, due to the wide number of applications to which they could contribute, which does not cease to increase~\cite{Klitzing1980,Shukai2020,Gilbert2021}.

A particular branch of topological systems that has rapidly evolved is the one of periodically driven setups or Floquet phases~\cite{lindner_floquet_2011,Kitagawa2011,AlvaroFloquet,Delplace2013,Grushin2014,Perez-Piskinow2014,mipaper2}.
Initially, they were in the spotlight for their external control, which allows to simulate static topological phases that would be extremely hard to realize by other means. However, the discovery of anomalous Floquet topological phases~\cite{Rudner2013,Gomez-Leon2014,MoraisAF,Rudner2020,Gomez-LeonAnomalous2023}, a unique phenomena of periodically driven systems where a system with topologically trivial bands displays topologically protected edge states, renewed the interest in the topology of Floquet systems.

Simultaneously, the research in cavity-QED (c-QED) materials is booming due to recent experimental advances that allow to explore new regimes of light-matter interaction~\cite{Forn2010,Niemczyk2010,Yoshihara2017,frisk_kockum_ultrastrong_2019}. In this case, the material couples to quantized light and forms an isolated many-body state with properties dictated by their mutual influence.
Here, the interest is not only to understand in more detail the fundamental interactions between light and matter, but also from an applied perspective, to tame these interactions for their use in quantum technologies~\cite{Romero2012,cavityquantummat}.

Recently, a link between c-QED materials and Floquet physics has unfolded. In particular, it has been demonstrated that the semi-classical limit of c-QED materials features aspects of Floquet physics, such as the band renormalization by Bessel functions~\cite{quantumtoclassical}.
Attempts to further understand this emergence of an effective Floquet description in the semi-classical limit of c-QED materials has given birth to Quantum Floquet engineering~\cite{Eckstein2020,Eckstein2022,Dmytruk2022,floquetEngKennes,TopologyDet}.
\begin{figure}
    \centering
    \includegraphics[width=\columnwidth]{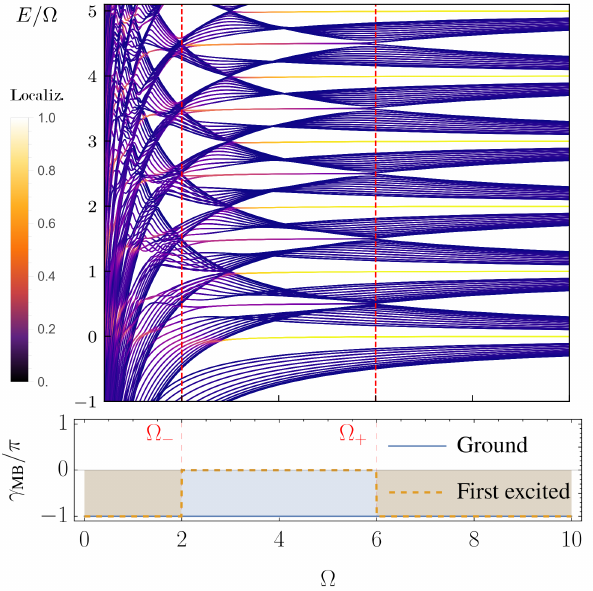}
    \caption{Top: Many-body spectrum of a topological SSH chain ($J^\prime=2$) interacting with a cavity, as a function of its frequency for $g=0.35$. The interaction produces many-body gaps with edge states for the range $\Omega\in [\Omega_+, \Omega_-]$. Color code indicates localization at the edge. Bottom: Many-body Zak phase for the ground and first excited states. All plots in units of $J$.}
    \label{fig:Fig1}
\end{figure}

Intuitively, their relation can be understood from the point of view that, if a cavity field in its steady-state is traced-out, the matter effectively couples to a time-periodic drive with frequency set by the cavity. However, for this to be true both, the backaction from the matter onto the cavity field and the light-matter correlations must be negligible. This defines the semi-classical limit of c-QED materials to effectively obtain the physics of Floquet systems~\cite{quantumtoclassical,perezgonzalez2023lightmatter}.

This relation is not a mere curiosity and has profound implications. On the one hand, it shows that there is a microscopic mechanism by which the discrete time-translation symmetry of Floquet systems $H(t)=H(t+T)$, effectively emerges from a many-body, isolated quantum system. On the other hand, it also provides a bridge between widely different conserved quantities: the energies of the c-QED material and the quasienergies of the Floquet system, which notably, are defined modulo $2\pi$ only.

In this work we shed light into these questions by asking: \emph{what is the many-body origin of anomalous Floquet topological phases?}.
Notice that anomalous Floquet phases strongly depend on the hypothesis of discrete time periodicity. It is because of this symmetry that their topological classification goes beyond that of static systems~\cite{Roy2017}.

We show that anomalous Floquet topological phases emerge in c-QED materials when the cavity mode is resonant with the band transitions of a topological system. If this resonant interaction has the right symmetries, it opens additional many-body gaps that contain edge states, and in contrast with the original ones, they are made of entangled light-matter excitations.
We find that despite the existence of these additional edge states, the many-body topological invariant vanishes, invalidating the bulk-boundary correspondence, in analogy with the anomalous topology of Floquet phases.
To resolve this, we demonstrate that it is possible to assign two topological invariants to the system and through them, establish a bulk-boundary correspondence that perfectly predicts the existence of edge states.
Furthermore, the existence of these pair of invariants allows us to identify the $\pi$-gap of Floquet systems with the many-body gaps opened by this resonant interaction, establishing a link between many-body and periodically driven systems.
\section{Results}
The Su-Schrieffer-Heeger (SSH) model has been a canonical model of a topological insulator in one dimension for a long time~\cite{SSH0,SSH1,Jackiw1976,mipaper1}.
Not only in the static case its topology has been thoroughly studied, but also its periodically driven version provides a deep understanding of Floquet physics~\cite{Gomez-Leon2013}.
For example, when coupled to light via the Peierls substitution, one can easily show the control of the stroboscopic Hamiltonian at high frequency, with the possibility to externally tune the presence of topological edge states~\cite{mipaper2}.
Although in that case lowering the frequency has an impact on the edge states,  anomalous topology emerges only for particular driving protocols that preserve chiral symmetry and are resonant with the system~\cite{Balabanov2017}. In that case, topological edge states populate all the unequivalent quasienergy gaps at the same time, while the winding number of the quasienergy bands remains trivial~\cite{Gomez-LeonAnomalous2023,Qingqing2019}.

This coexistence of topological edge states with topologically trivial Floquet bands is a hallmark of Floquet phases without a static analog. For this reason, if Floquet phases are to arise from c-QED materials in a semi-classical limit, it is important to understand their quantum many-body origin.
For that, we study a SSH chain interacting with a single mode cavity via a coupling term that preserves chiral symmetry:
\begin{equation}
    H=\Omega d^{\dagger}d+\sum_{j=1}^{N} \left( J_1[d] b_{j}^{\dagger}a_{j}+J_2[d]a_{j+1}^{\dagger}b_{j}+\text{h.c.} \right), \label{eq:Hamiltonian1}
\end{equation}
where we have defined the state-dependent hopping $J_{1}\left[d\right]\equiv J+g\left(d+d^{\dagger}\right)$ and $J_{2}\left[d\right]\equiv J^{\prime}-g\left(d+d^{\dagger}\right)$.
The first term in Eq.~\eqref{eq:Hamiltonian1} describes a cavity with frequency $\Omega$ and photon operators $d$ and $d^\dagger$, while the second term describes the hopping of spinless fermions between sites A and B of the chain with operators $a_j$, $a_j^\dagger$, $b_j$ and $b_j^\dagger$.
Importantly, the hopping in the chain is not only dimerized with $J$ and $J^\prime$, but also modulated by the absorption/emission of photons, with opposite sign for the intra- and inter-dimer hopping.

As Eq.~\eqref{eq:Hamiltonian1} describes a complex many-body problem, let us first introduce the basic properties of the unperturbed SSH chain Hamiltonian $H_\text{SSH}=H(g=0)$.
For Periodic Boundary Conditions (PBC) it can be exactly diagonalized and results in two eigenstates, $| \varphi_{\pm} (k) \rangle$, with energies:
\begin{equation}
    E_{\pm}\left(k\right)=\pm\sqrt{J^{2}+J^{\prime2}+2JJ^{\prime}\cos\left(k\right)}. \label{eq:Energy1}
\end{equation}
As the Hamiltonian has chiral symmetry: $\sigma_z H_\text{SSH} \sigma_z=-H_\text{SSH}$, the topology can be characterized by a winding number:
\begin{equation}
    \nu_0 =\int_{-\pi}^{\pi} \frac{dk}{4\pi i} \text{tr} \left[ \sigma_z H_\text{SSH}^{-1} \partial_k H_\text{SSH} \right], \label{eq:winding1}
\end{equation}
or equivalently by the Zak phase of the eigenvectors $\gamma_0=\pi\nu_0$~\cite{Berry1984,Zak1989,Atala2013}.
In particular, for the unperturbed SSH chain $\nu_0= -\Theta ( J^\prime/J -1)$, predicting the existence of edge states for $J^\prime > J$.

When $g\neq 0$, Eq.~\eqref{eq:Hamiltonian1} can be exactly diagonalized for chains of reasonable size.
In Fig.~\ref{fig:Fig1} we plot the energy levels as a function of the cavity frequency for a topological chain ($J<J^\prime$).
It shows that as the cavity becomes resonant with the chain, additional edge states emerge within newly formed many-body gaps.

To try to topologically characterize these additional edge states one can extend the Zak phase analysis of the SSH chain to the \emph{many-body} Zak phase $\gamma_\text{MB}$~\cite{Grusdt2019}, with $\nu_\text{MB}=\gamma_\text{MB}/\pi$ the corresponding winding number. Its calculation is shown in Fig.~\ref{fig:Fig1} for the ground and first excited states, and in particular, one finds that the first excited state becomes trivial when the resonance is reached.
This situation reminds of the anomalous Floquet topological phase, where with the emergence of edge states in the $\pi$-gap, the Floquet band invariant becomes trivial~\cite{Balabanov2017,Gomez-LeonAnomalous2023,Rudner2013,Rudner2020}.
However, notice that the system is isolated and many-body in our case.
We will now demonstrate that this many-body phase with resonantly coupled photons and fermions is not topologically trivial.

In order to describe the topology, it is enough to restrict our analysis to the weak coupling regime of Eq.~\eqref{eq:Hamiltonian1}. This is a good approximation if we are interested in changes in the topology produced by a near resonant cavity. The reason is that weak coupling ensures that, far from the resonance, the topology of the SSH chain is approximately unaffected by the cavity. Then, as the cavity frequency is reduced and matches the chain energy gap, only the resonance can change the topology, because it corresponds to a non-perturbative mechanism. 
On top of this, different Hamiltonians in the weak coupling regime can lead to identical effective low-energy interactions with the cavity, which makes our model a valid description for various experimental realizations that will be discussed below.
\begin{figure*}
    \centering
    \includegraphics[width=.9\textwidth]{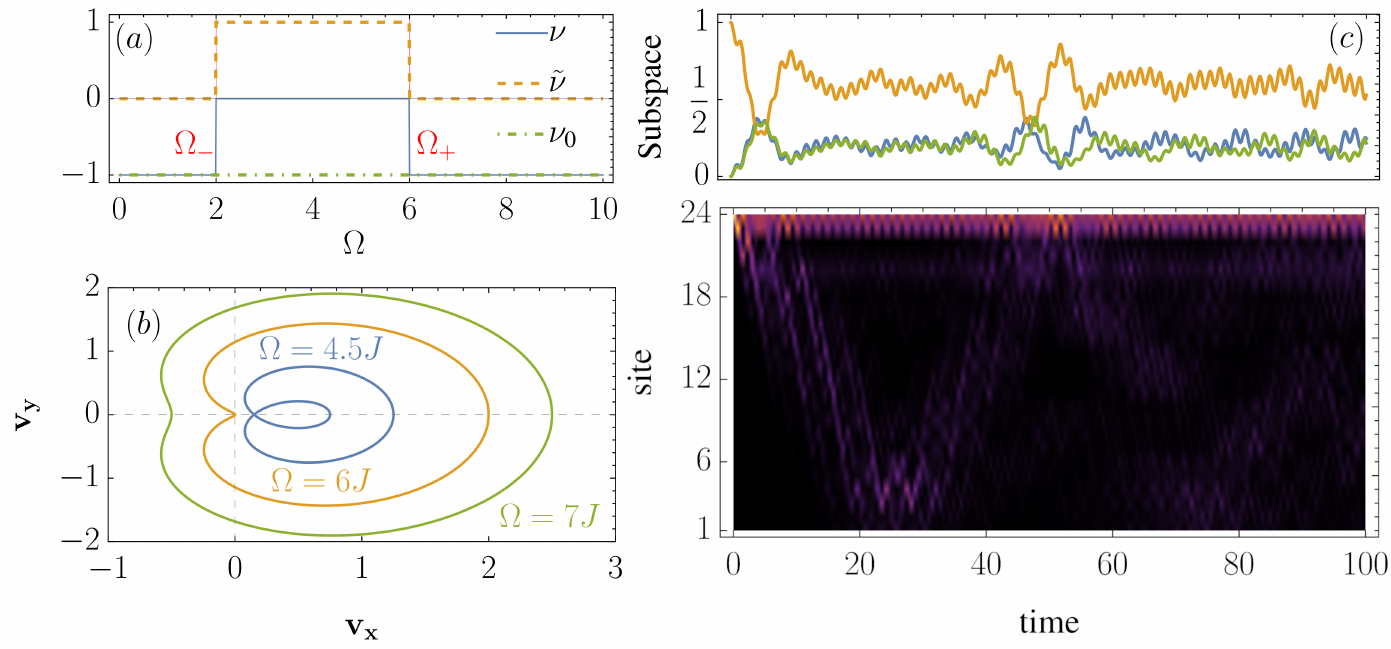}
    \caption{$(a)$ Winding number as a function of cavity frequency. Contribution from the RWA Hamiltonian $\nu$ (solid), from the resonance $\tilde{\nu}$ (dashed), and from the the unperturbed SSH chain $\nu_0$ (dot-dashed). Gap closures at resonances $\Omega_{\pm}$ in vertical dashed lines. $(b)$ Bloch vector trajectories for high ($\Omega=7$), resonant ($\Omega=6=\Omega_+$) and below resonant frequency ($\Omega=4.5$). The other parameters are $J^\prime=2$, $g=0.1$ and $n=0$. $(c)$ Dynamics for an initial state localized at the edge of a chain with $J^\prime=J/2$ and $L=24$, coupled to a cavity with $\Omega=5J^\prime$ and $n=5$ photons. Top panel shows the occupation of the photon subspaces $| n \rangle$ (yellow) and $| n \pm 1 \rangle$ (blue and green), while bottom panel shows the dynamics of a particle initially at the edge.}
    \label{fig:Fig2}
\end{figure*}

To proceed, first notice that if the cavity frequency is much larger than the bandwidth, $2|J+J^\prime|$, the bands in Eq.~\eqref{eq:Energy1} are good approximations in the weak coupling regime. Therefore, it is reasonable to re-write the interaction term in this basis and retain only the terms that can lead to resonances. This Rotating Wave Approximation (RWA) leads to a Jaynes-Cummings model with a momentum dependence (details in the Appendix). Finally, noticing that the total number of excitations is conserved, we can write the Hamiltonian in a block diagonal form with basis elements: $\{ |\varphi_+, n\rangle, |\varphi_-, n+1\rangle \}$:
\begin{equation}
    \tilde{H}\left(n,k\right) = \left(\begin{array}{cc}
n\Omega+E_{+}\left(k\right) & -ig\Gamma\left(k\right)\sqrt{n+1}\\
ig\Gamma\left(k\right)\sqrt{n+1} & \left(n+1\right)\Omega+E_{-}\left(k\right)
\end{array}\right) \label{eq:Hamiltonian2},
\end{equation}
where we have defined $\Gamma (k) \equiv \left(J+J^{\prime}\right) \sin\left(k\right)/E_+(k)$.
Equation~\eqref{eq:Hamiltonian2} can be directly diagonalized and results in the following bands:
\begin{align}
    \epsilon_{\pm}\left(k\right)=& \Omega\left(n+\frac{1}{2}\right) \nonumber \\
    &\pm\sqrt{\left[E_{+}\left(k\right)-\frac{\Omega}{2}\right]^{2}+g^{2}\left(n+1\right)\Gamma\left(k\right)^{2}},
\end{align}
with a shift $n \Omega$ for each subspace, a $\Omega$-dependent gap which is reduced as the resonance $\Omega=2E_+(k)$ is approached, and an anti-crossing due to the transverse light-matter coupling, proportional to $g^2 (n+1) \Gamma (k)$.

Crucially, the anti-crossing becomes an exact crossing at the high symmetry points $k_0 = m \pi$, for $m\in \mathbb{Z}$. Hence, we can predict the frequencies at which the gap exactly closes due to the resonance. 
We find that for $\Omega_{\pm} = 2E_+(k_0) = 2| J \pm J^{\prime} |$, the gap closes at $k= 0,\pi$, respectively.
Red vertical lines in Fig.~\ref{fig:Fig1} indicate these values of the frequency, and one can see that they perfectly predict the appearance of additional edge states in the many-body gaps induced by the resonant coupling between cavity and system.

To gain further intuition, we also re-write the RWA Hamiltonian for each block in the original basis:
\begin{equation}
    H(n,k) = \left(\begin{array}{cc}
\left(n+\frac{1}{2}\right)\Omega & E_{+}\frac{E_{+}-\frac{\Omega}{2}-ig\sqrt{n+1}\Gamma}{J+J^{\prime}e^{-ik}}\\
E_{+}\frac{E_{+}-\frac{\Omega}{2}+ig\sqrt{n+1}\Gamma}{J+J^{\prime}e^{ik}} & \left(n+\frac{1}{2}\right)\Omega
\end{array}\right). \label{eq:Hamiltonian3}
\end{equation}
Clearly, it retains a chiral form in each subspace, and in the high frequency regime $\Omega \gg J,J^\prime$, describes copies of the SSH chain Hamiltonian, correctly predicting the topology for a largely detuned cavity.
However, the off-diagonal elements now describe the hopping of polaritons, which near the resonance can strongly affect the topology due to the dominance of the term proportional to $g$.

Because of the chiral form of $H(n,k)$, the winding number $\nu$ can be used for the topological characterization. Its calculation is equivalent to that of Eq.~\eqref{eq:winding1}, with $H(n,k)$ instead of $H_\text{SSH}$, and leads to Fig.~\ref{fig:Fig2}, panel $(a)$ (solid line).
As predicted, at high frequency the topology is identical to that of the unperturbed SSH chain (dot-dashed line). However, at $\Omega_+=2|J+J^\prime|$, the winding number has an additional contribution with opposite sign that predicts a trivial topological phase.
Notice the perfect agreement with the invariant from the many-body Zak phase in Fig.~\ref{fig:Fig1}, for the first excited state. This confirms that the RWA Hamiltonian correctly describes the relevant physics for the topological changes.
Furthermore, our effective Hamiltonian in Eq.~\eqref{eq:Hamiltonian3} explains why the ground state topology remains unaffected: the state $|\varphi_-,n=0\rangle$ is completely decoupled from the rest.

The change in the invariant can be understood graphically. In Fig.~\ref{fig:Fig2}, panel $(b)$, we plot the trajectory of the Bloch vector $\mathbf{v}$ in $H(n,k)=\mathbf{v}\cdot \boldsymbol{\sigma}$, as $k$ is varied across the First Brillouin Zone (FBZ). It shows that at high frequency, the trajectory encloses the origin counterclockwise and results in $\nu=-1$, as expected for the SSH chain. However, after the resonance the Bloch vector winds twice, but does not enclose the origin and results in $\nu=0$~\cite{MartinMoreno2019}.

To understand the origin of the additional contribution to $\nu$ at resonance, we explicitly calculate the Zak phase of the many-body eigenstates:
\begin{equation}
    |\psi_{\mu}\left(k,n\right)\rangle = \alpha_{\mu}\left(k,n\right)|n,\varphi_{+}\left(k\right)\rangle + \beta_{\mu}\left(k,n\right)|n+1,\varphi_{-}\left(k\right)\rangle \label{eq:Eigenvector1}
\end{equation}
with $\alpha_{\pm}$ and $\beta_{\pm}$ the components of the eigenvectors that diagonalize Eq.~\eqref{eq:Hamiltonian2} and $| \varphi_\pm  \rangle $ the eigenvectors that diagonalize the unperturbed SSH chain.
The Berry connection is:
\begin{align}
    \mathcal{A}_{\mu,\mu} =& i\langle\psi_{\mu}\left(k,n\right)|\partial_{k}\psi_{\mu}\left(k,n\right)\rangle \nonumber \\
    =& i\alpha_{\mu}^{*}\left(\partial_{k}\alpha_{\mu}\right)+i\beta_{\mu}^{*}\left(\partial_{k}\beta_{\mu}\right)+\mathcal{A}_0 , \label{eq:Connection1}
\end{align}
where $\mathcal{A}_0=\langle \varphi_{\pm} (k) | i\partial_k \varphi_{\pm}(k)\rangle$ is the Berry connection for the unperturbed SSH chain and we have used that $\left|\alpha_{\mu}\right|^{2}+\left|\beta_{\mu}\right|^{2}=1$.
This shows that the connection separates in two contributions, one coming from the unperturbed SSH chain, and another coming from the resonant interaction. The latter is a purely many-body effect that entangles each band with a subspace with a different number of photons.
The Zak phase is obtained by integrating the Berry connection over the FBZ:
\begin{equation}
    \gamma = \int_{-\pi}^\pi \mathcal{A}_{\mu,\mu} dk = \pi \nu = \pi (\tilde{\nu}+\nu_0) \label{eq:Zak1},
\end{equation}
and as it is quantized in chiral systems and $\mathcal{A}_0$ is independent of the interaction with the cavity, the new contribution from light-matter interactions $\tilde{\nu}$, must also be quantized.
This is shown in Fig.~\ref{fig:Fig2}, panel $(a)$, where we plot all contributions. It can be seen that, in the resonant regime $\Omega\in [\Omega_+,\Omega_-]$, there is a one-to-one correspondence between the change in the total invariant $\nu$ and the contribution from the resonance $\tilde{\nu}$, which in combination with $\nu_0$, leads to the incorrect conclusion that the system is topologically trivial.

From these findings we can conclude that the winding number of $H(k,n)$ is not a good topological invariant to characterize the presence of topological edge states. Instead, it must be separated in two contributions, one corresponding to the winding number of the unperturbed SSH chain $\nu_0$ and another to the winding number $\tilde{\nu}$ corresponding to the RWA Hamiltonian in the eigenstates basis, Eq.~\eqref{eq:Hamiltonian2}. This invariant is quantized and completely predicts the appearance of edge states in the additional many-body gaps due to the resonant interaction.

To visualize the many-body origin of the edge states induced by the interaction, we plot in Fig.~\ref{fig:Fig2} panel $(c)$, the dynamics of a particle initially at the edge of a chain with $J^\prime<J$ (lower panel). This configuration ensures the absence of edge states due to the topology of the isolated chain. The particle remains largely localized over time, stating the presence of edge states, but more importantly, the upper panel shows the occupation of the photon subspaces $| n \rangle$ and $|n \pm 1 \rangle$, reflecting the light-matter correlations of the analytical solution. In particular, the cavity is initially prepared in $|n=5\rangle$ (yellow), and the interaction leads to an oscillation with $|n\pm 1\rangle$ (blue and green), in correlation with the localized particle dynamics (additional cases in the Appendix).
\section{Conclusions}
In summary, our results provide a bridge between isolated many-body systems and periodically driven ones.
We have shown that anomalous Floquet topological phases can emerge in the semi-classical limit of c-QED materials and have described their quantum many-body origin.
Far from the semi-classical limit they also display edge states and trivial bands, but their physical properties are very different due to the presence of backaction and light-matter correlations.
This confirms that an effective, discrete time-translation symmetry emerges in the matter sector due to the coupling to cavity photons, as otherwise anomalous topological phases could not be present.

The required ingredient to create this many-body phase in c-QED materials is a cavity resonant with a topological system. However, the interaction term also needs to have the right symmetry, to produce an exact crossing between different Fock subspaces.

To confirm the topological origin of the many-body phase we have identified two topological invariants, $\nu_0$ and $\tilde{\nu}$, which predict the presence of edge states in the single-particle and many-body gaps, respectively. One is the standard winding number of non-interacting one-dimensional chiral systems, while the other is a winding number that captures the resonant many-body interaction.
We have shown that although their sum vanishes, it is their independent value that matters and defines a many-body invariant for the system, which establishes the correct bulk-to-boundary correspondence.
Importantly, our result shows a link between the unequivalent $0$ and $\pi$ gaps of Floquet systems, and the single particle and many-body gaps induced by the resonant interaction, respectively.

Our findings could be experimentally verified in different platforms. For example, in ion traps it should be possible to engineer the required interaction using an additional phononic degree of freedom, as dimerization has been already achieved~\cite{Nevado2017}. Importantly, the phonons would need to have the right frequency, for their interaction with the chain to become resonant.
Another possibility would be the use of superconducting circuits~\cite{SCCircuitsProposal1}, where the coupling to the cavity can be simulated by additional waveguides.
\begin{acknowledgments}
We acknowledge support from the European Union’s Horizon2020 research and innovation program under Grant Agreement No.899354(SuperQuLAN), the Proyecto Sinergico CAM 2020 Y2020/TCS-6545 (NanoQuCoCM) and from CSIC Interdisciplinary Thematic Platform (PTI+) on Quantum Technologies (PTI-QTEP+).
\end{acknowledgments}
\bibliography{BibliographyQFE}
\onecolumngrid
\appendix
\section{Derivation of the effective Hamiltonian}
The effective Hamiltonian can be directly derived from the original Hamiltonian:
\begin{equation}
    H=\Omega d^{\dagger}d+\sum_{j=1}^{N} \left( J_1[d] b_{j}^{\dagger}a_{j}+J_2[d]a_{j+1}^{\dagger}b_{j}+\text{h.c.} \right)
\end{equation}
where we have defined the state-dependent hopping $J_{1}\left[d\right]\equiv J+g\left(d+d^{\dagger}\right)$ and $J_{2}\left[d\right]\equiv J^{\prime}-g\left(d+d^{\dagger}\right)$.

The first step is to find the eigenvectors $| \varphi_\pm (k) \rangle$ of the unperturbed SSH chain Hamiltonian for PBC:
\begin{equation}
    H_{\text{SSH}} = \sum_{k}\left(a_{k}^{\dagger},b_{k}^{\dagger}\right) \left(\begin{array}{cc}
0 & J+J^{\prime}e^{ik}\\
J+J^{\prime}e^{-ik} & 0
\end{array}\right)\left(\begin{array}{c}
a_{k}\\
b_{k}
\end{array}\right)
\end{equation}
The corresponding eigenvalues are $E_\pm (k) = \pm\sqrt{J^{2}+J^{\prime2}+2JJ^{\prime}\cos\left(k\right)}$. Then, rewritting the full Hamiltonian in this basis for the fermionic part, we arrive at:
\begin{equation}
    \tilde{H} = \Omega d^{\dagger}d + \sum_{k,\nu=\pm}E_{\nu}\left(k\right) c^\dagger_{k,\nu} c_{k,\nu} + g\left(d+d^{\dagger}\right)\sum_{k}\sum_{\mu,\nu=\pm}V_{\mu,\nu}\left(k\right) c^\dagger_{k,\mu} c_{k,\nu} \label{eq:Hamiltonian1-1}
\end{equation}
where $c^\dagger_{k,\nu}$ and $c_{k,\nu}$ are the creation/annihilation operators of fermionic eigenstates in the unperturbed SSH chain, which are linear combinations of $a_k$ and $b_k$ operators, with $k$-dependent coefficients.
In particular, the matrix elements of the coupling between the two systems are given by the following expressions:
\begin{align}
    V_{+,+}\left(k\right)&=\frac{\left(1-\cos\left(k\right)\right)\left(J-J^{\prime}\right)}{\sqrt{J^{2}+J^{\prime2}+2JJ^{\prime}\cos\left(k\right)}}\\V_{+,-}\left(k\right)&=\frac{-i\sin\left(k\right)\left(J+J^{\prime}\right)}{\sqrt{J^{2}+J^{\prime2}+2JJ^{\prime}\cos\left(k\right)}}=-i\Gamma\left(k\right)\\V_{-,+}\left(k\right)&=\frac{i\sin\left(k\right)\left(J+J^{\prime}\right)}{\sqrt{J^{2}+J^{\prime2}+2JJ^{\prime}\cos\left(k\right)}}=i\Gamma\left(k\right)\\V_{-,-}\left(k\right)&=-\frac{\left(1-\cos\left(k\right)\right)\left(J-J^{\prime}\right)}{\sqrt{J^{2}+J^{\prime2}+2JJ^{\prime}\cos\left(k\right)}}
\end{align}
Equation~\eqref{eq:Hamiltonian1-1} still is the full Hamiltonian, but it can be truncated in this new basis. For that, notice that we are interested in the effect of a resonant interaction for weak coupling. Hence, as this is a non-perturbative process, we can just keep the rotating terms of the interaction and ignore the counter-rotating ones, which will enter as small perturbations and can be accounted for later, if necessary.
After the RWA, the Hamiltonian becomes:
\begin{equation}
    \tilde{H} = \Omega d^{\dagger}d + \sum_{k,\nu=\pm}E_{\nu}\left(k\right) c^\dagger_{k,\nu} c_{k,\nu} + i g\sum_{k} \Gamma\left(k\right)\left(d^{\dagger}c_{k,-}^{\dagger}c_{k,+}-dc_{k,+}^{\dagger}c_{k,-}\right) \label{eq:Hamiltonian2-1}
\end{equation}
This Hamiltonian is analogous to a Jaynes-Cummings Hamiltonian with a $k$-dependence, as it can be seen by just defining Pauli matrices $\sigma_k^{\pm}$ from the creation/destruction fermion operators.
As it conserves the total number of excitations, we can write it in block-diagonal matrix form in the basis $\{ |\varphi_{+},n\rangle,|\varphi_{-},n+1\rangle \}$, and find for each block:
\begin{equation}
    \tilde{H}\left(n,k\right) = \left(\begin{array}{cc}
n\Omega+E_{+}\left(k\right) & -ig\Gamma\left(k\right)\sqrt{n+1}\\
ig\Gamma\left(k\right)\sqrt{n+1} & \left(n+1\right)\Omega+E_{-}\left(k\right)
\end{array}\right)\label{eq:AnalyticalH}
\end{equation}
Finally, its diagonalization results in the eigenvalues found in the main text:
\begin{equation}
    \epsilon_{\pm}\left(k\right)=\Omega\left(n+\frac{1}{2}\right)\pm\sqrt{\left[E_{+}\left(k\right)-\frac{\Omega}{2}\right]^{2}+g^{2}\left(n+1\right)\Gamma\left(k\right)^{2}}
\end{equation}
For illustrative purposes, it is also interesting to rotate back to the original frame the RWA Hamiltonian. There, one can explicitly see the chiral symmetry of the model, once the counter-rotating terms have been removed and resonance effect has been taken into account. It takes the following form:
\begin{equation}
    H(n,k) \approx \left(\begin{array}{cc}
\left(n+\frac{1}{2}\right)\Omega & E_{+}\frac{E_{+}-\frac{\Omega}{2}-ig\sqrt{n+1}\Gamma}{J+J^{\prime}e^{-ik}}\\
E_{+}\frac{E_{+}-\frac{\Omega}{2}+ig\sqrt{n+1}\Gamma}{J+J^{\prime}e^{ik}} & \left(n+\frac{1}{2}\right)\Omega
\end{array}\right) \label{eq:EffectiveHLab1}
\end{equation}
\section{Topological invariants}
In this section we provide additional calculations of the topological invariants and Berry phases.

Starting from the simplest case of the unperturbed SSH chain, where chiral symmetry is given by a $\sigma_z$ operator, we can obtain the Berry connection from the eigenstates of $H_\text{SSH}$, $| \varphi_\mu \rangle$:
\begin{equation}
    \mathcal{A}_0 = \langle\varphi_{\mu}|i\partial_{k}|\varphi_{\mu}\rangle =-\frac{J^{\prime}\left[J\cos\left(k\right) +J^{\prime}\right]}{2\left[J^{2}+J^{\prime2}+2JJ^{\prime}\cos\left(k\right)\right]}
\end{equation}
Interestingly, the expression is identical for the two eigenstates. This is the reason why the calculation of the Zak phase, by integrating over the whole FBZ, results in the same values for both (we define $z\equiv e^{ik}$):
\begin{equation}
    \gamma_0 = -\oint\frac{dz}{4i}\frac{z^{2}+2\frac{J^{\prime}}{J}z+1}{z\left(z-z_{+}\right)\left(z-z_{-}\right)} = -\pi \Theta\left(1-J/J^{\prime}\right)
\end{equation}
being $\Theta(x)$ the Heaviside function and the integral in the complex plane over the unit circle. This demonstrates that for the SSH chain, the Zak phase is quantized in the topological phase, $J<J^\prime$.

\begin{figure}
    \centering
    \includegraphics[width=0.4\textwidth]{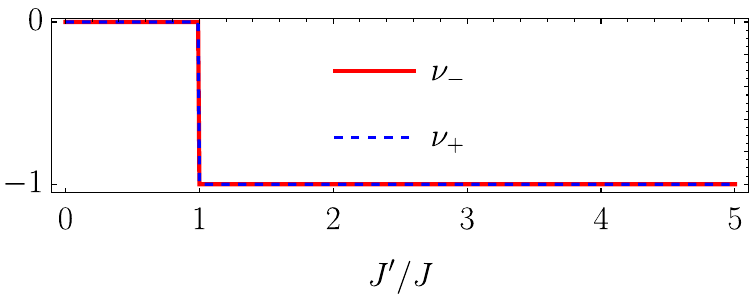}
    \caption{Winding number of each band of the unperturbed SSH model as a function of the ratio between intra- and inter-dimer hopping. The value is identical for the two bands.}
    \label{fig:Winding1}
\end{figure}

Because the system has chiral symmetry, it is also possible to directly calculate the winding number from the Hamiltonian:
\begin{equation}
    \nu_0 = \int_{-\pi}^{\pi}\text{tr}\left\{ \sigma_{z}H_{\text{SSH}}^{-1}\partial_{k}H_{\text{SSH}}\right\} \frac{dk}{4\pi i}= - \Theta\left(1-J/J^{\prime}\right), \label{eq:Winding0}
\end{equation}
which coincides with the expression from the Zak phase, divided by $\pi$. This particular feature allows to relate the quantized Zak phase with the value of the winding number via the relation $\gamma_0 = \pi \nu_0$.
Its value is shown in Fig.~\ref{fig:Winding1} as a function of the ratio between intra- and inter-dimer hopping.

\begin{figure}
    \centering
    \includegraphics[width=0.5\textwidth]{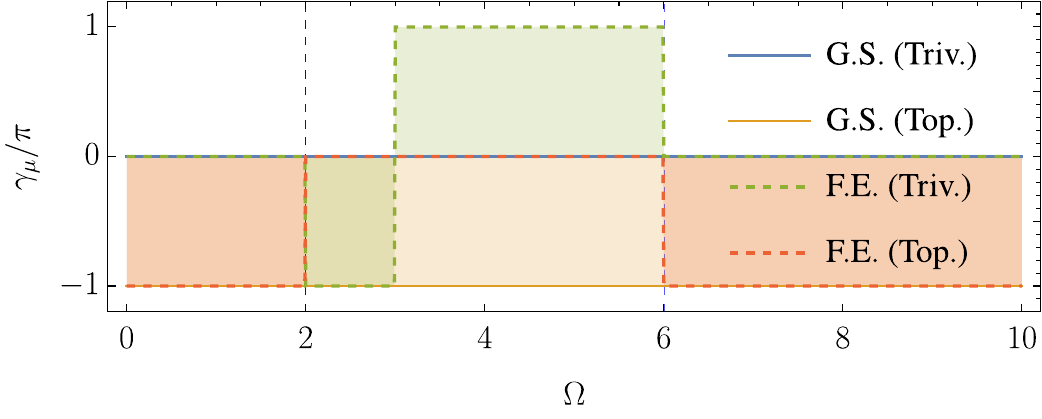}
    \caption{Many-body Zak phase in units of $\pi$ for the ground (solid) and first excited states (dashed), as a function of the cavity frequency. We have considered two types of configurations for the unperturbed chain: topological ($J>J^\prime$) and trivial ($J<J^\prime$). In particular the parameters are $g=0.1$, $J=2$ and $J^\prime=1$ for the trivial phase, while for the topological phase the hopping is swapped to $J=1$ and $J^\prime = 2$. This allows to keep identical bandwidth for both configurations. Dashed blue gird lines indicate the gap closure points due to resonances.}
    \label{fig:Winding2}
\end{figure}

This relation between the Zak phase and the winding number can be extended to the many-body Zak phase, if the many-body system is also chiral, as in our present case.
To calculate the many-body Zak phase we consider a momentum space discretization and calculate the infinitesimal rotation of many-body eigenstates, $| \Psi_\mu (k) \rangle$, as they are parallel transported along the FBZ. Summing over all contributions we arrive at:
\begin{equation}
    \gamma_\mu = -\sum_{k = -\pi}^{\pi - \delta k} \text{Im} \left\{ \log \left[ \langle \Psi_\mu (k) | \Psi_\mu (k+\delta k)\rangle \right] \right\},
\end{equation}
and define $\nu_\mu = \gamma_\mu/\pi$.
Notice that now the index $\mu$ of the many-body eigenstates runs over all many-body eigenstates.
In Fig.~\ref{fig:Winding2} we plot the value of the many-body Zak phase, for the ground and first excited state, as a function of cavity frequency.
Notice that the Zak phase for the ground state does not change with the frequency, because it is decoupled from the energy levels above it, while the first excited changes at the resonance, as predicted from the theory.
On top of this we can see that the first excited for the trivial phase changes sign at $\Omega=3$, which coincides with the closure of the single-particle gap in the spectrum. This is not captured with the analytical expression in the main text, because it requires to go beyond weak coupling.

\begin{figure}
    \centering
    \includegraphics[width=0.45\textwidth]{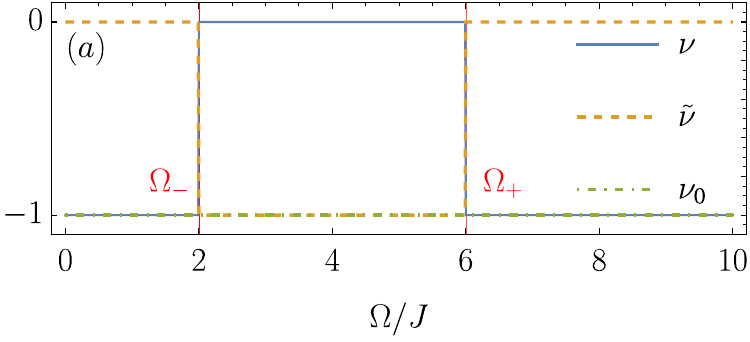}
    \caption{Winding numbers from the different contributions to the RWA Hamiltonian.}
    \label{fig:Winding-RWA}
\end{figure}

To complete the description of the topological phases, and in particular of the phase with many-body edge states and vanishing winding number, we focus on the analytically solvable model of Eq.~\eqref{eq:AnalyticalH}. The Hamiltonian is in the basis of eigenstates of $H_\text{SSH}$ and because it is chiral, its winding number for each subspace can be calculated with an identical formula as that of Eq.~\eqref{eq:Winding0}. Notice however that in order to use the same chiral operator, $\sigma_z$, one must perform a rotation of the Hamiltonian. Similarly, from its eigenstates one can calculate the Zak phase analytically, or numerically. All this gives the same result for $\tilde{\nu}$, shown in Fig.~\ref{fig:Winding-RWA} and in the main text. 
It shows that the resonance contributes with a change to the winding number in one unit, because under resonance the eigenstates rotate an additional $\pi$ angle when parallel transported along the FBZ. Hence, as the SSH chain initially is in a topological phase (see the value of $\nu_0$ in Fig.~\ref{fig:Winding-RWA}), the total, many-body Zak phase rotates a multiple of $2\pi$ and it looks as if the eigenstates are trivially transported. 
However, our result shows that the bulk-to-edge correspondence is only captured if we separate the single particle and the many-body contribution from the resonance.

To confirm this Fig.~\ref{fig:Winding-RWA} also shows the value of the winding number from the RWA Hamiltonian in the original basis (Eq.~\eqref{eq:EffectiveHLab1}), $\nu$, which contains the contribution from the unperturbed SSH chain as well and perfectly agrees with the many-body Zak phase that gives a trivial value in the resonant regime.

To explicitly see the separation between the two contributions one just needs to calculate the Berry connection for the RWA Hamiltonian. As its eigenstates are of the form:
\begin{equation}
    |\psi_{\mu}\left(k,n\right)\rangle=\alpha_{\mu}\left(k,n\right)|n\rangle|\varphi_{+}\left(k\right)\rangle+\beta_{\mu}\left(k,n\right)|n+1\rangle|\varphi_{-}\left(k\right)\rangle
\end{equation}
with $\alpha_\mu(k,n)$ and $\beta_\mu(k,n)$ the coefficients of the eigenstates.
Its Berry connection results in the following two contributions:
\begin{equation}
    \mathcal{A}_{\mu,\mu} =	i\alpha_{\mu}^{*}\left(\partial_{k}\alpha_{\mu}\right)+i\beta_{\mu}^{*}\left(\partial_{k}\beta_{\mu}\right)+\mathcal{A}_0
\end{equation}
where the last term is the unperturbed Berry connection, and the first two terms describe the contribution from the hybridization between light and matter. As the integral over the FBZ is the Zak phase and must be quantized, this means that the contribution from the resonance must also be quantized, as we can write the total Zak phase as:
\begin{equation}
    \gamma=\gamma_0+\tilde{\gamma}
\end{equation}
or similarly for the winding number:
\begin{equation}
    \nu=\nu_0+\tilde{\nu}
\end{equation}
\section{Many-body dynamics}
In this section we provide additional plots for the many-body dynamics in the presence of different types of edge states.
This is interesting to understand the fundamental differences between the original SSH chain edge states, which are present for a highly detuned cavity and are approximately decoupled from the photons (i.e., single particle behavior), from the many-body ones, which are strongly correlated with the photons and oscillate in time.

\begin{figure}
    \centering
    \includegraphics[width=0.32\textwidth]{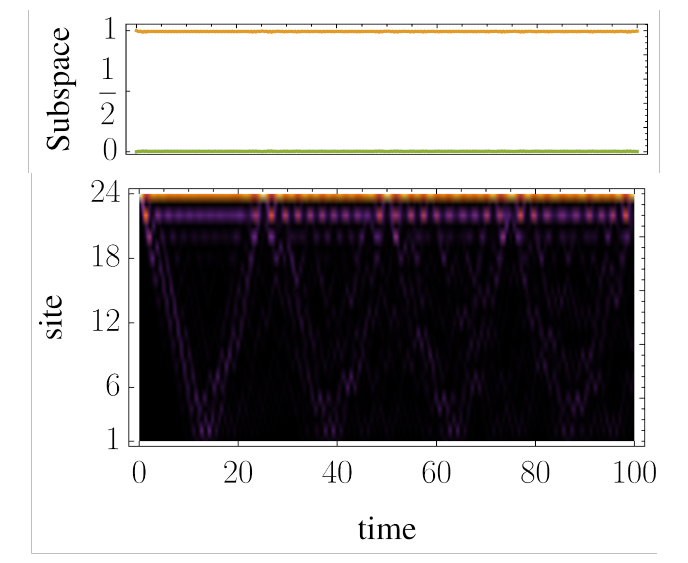}
    \includegraphics[width=0.32\textwidth]{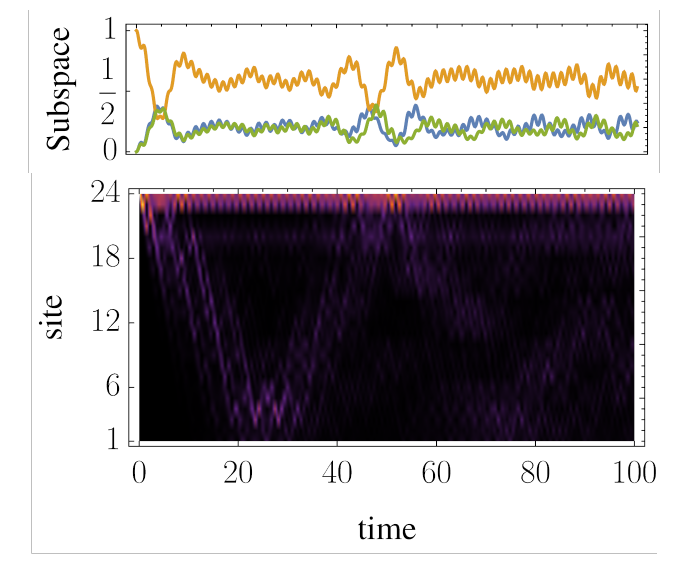}
    \includegraphics[width=0.32\textwidth]{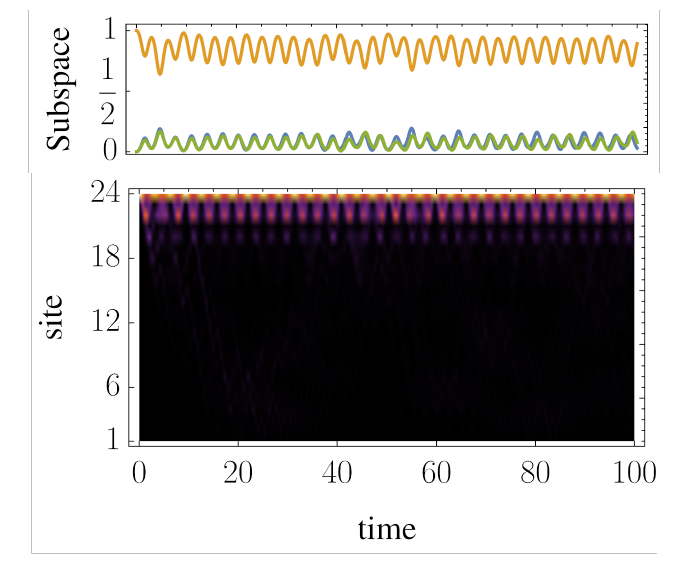}
    \caption{Bottom row shows the exact diagonalization dynamics of a fermion initially located at site $L=24$ of the chain, for a cavity with $n=5$ photons. Top row shows the occupation of Fock subspaces with $n=4,5$ and $6$.
    Left column considers a topological chain $J^\prime=2J$ for a largely detuned cavity with $\Omega=10$. Middle column considers a resonant cavity with $\Omega=4$ for an initially trivial chain, $J=2J^\prime$. Right column considers a resonant cavity with $\Omega=4$ for an initially topological chain, $J^\prime=2J$. In all plots the light-matter coupling is set to $g=0.1$.}
    \label{fig:Dynamics1}
\end{figure}

In Fig.~\ref{fig:Dynamics1} we show the matter and photon dynamics for an initially localized fermion at one edge of the chain, $x=L=24$, and a cavity with $n=5$ photons. 
First, the left column considers the case of a largely detuned cavity, coupled to a topological SSH chain ($J^\prime=2J$). It shows an edge state with very similar properties to those of the unperturbed SSH chain. On top, the photon occupation remains almost constant, indicating that the light and matter degrees of freedom are weakly hybridized.
The middle column shows the opposite situation, where an initially trivial SSH chain is in resonance with the cavity photons. In that configuration edge states in the many-body gaps are present, and their light-matter correlations are significant. This is shown in the dynamics of the different photon subspaces, where the cavity coherently oscillates between $n=4,5$ and $6$ photon states, as predicted from theory. The dynamics corresponds to the absorption/emission of photons from the fermion at the edge, which remains spatially localized.
Finally, the right column shows the case with vanishing total winding number, $\nu=0$, which displays both types of edge states (those coming from the unperturbed SSH chain and the ones coming from the resonant interaction with the cavity). In that case the fermion still is localized at the edge and light-matter correlations are a combination of the two previous situations.
\end{document}